\documentclass[twocolumn,english,aps,superscriptaddress]{revtex4-2}
\usepackage[T1]{fontenc}
\usepackage[utf8]{inputenc}
\usepackage{amsmath}
\usepackage{graphicx}
\usepackage{esint}

\makeatletter
\usepackage{amsthm}
\usepackage{amsfonts}
\usepackage{bbm}
\usepackage{epsfig}
\usepackage{babel}
\usepackage{color}
\usepackage{framed}
\usepackage{changes}
\usepackage{float}
\usepackage{array}

\usepackage{hyperref}
\hypersetup{colorlinks=true,citecolor=red,linkcolor=blue}  
\usepackage[utf8]{inputenc}

\makeatother

\usepackage{babel}
\begin{document}
\title{Radial Oscillations of Scalar Hair in Black Hole Bombs }
\author{Lang Zhao}
\email{202320130090@mail.scut.edu.cn}

\affiliation{\textit{School of Physics and Optoelectronics, South China University
of Technology, Guangzhou 510641, China}}
\author{Lin Chen}
\email{Corresponding author: linchen91@scut.edu.cn }

\affiliation{\textit{School of Physics and Optoelectronics, South China University
of Technology, Guangzhou 510641, China}}
\author{Cheng-Yong Zhang}
\email{Corresponding author: zhangcy@email.jnu.edu.cn}

\affiliation{\textit{Department of Physics and Siyuan Laboratory, Jinan University,
Guangzhou 510632, China}}
\begin{abstract}
Recent research has revealed a novel nonlinear mechanism, distinct
from the linear superradiant instability, which triggers the black
hole bomb phenomenon. Introducing a massive complex scalar field with
nonlinear self-interactions drives the Reissner-Nordström black hole
to shed substantial energy, thereby triggering a black hole bomb.
Radial oscillations in the scalar hair profile are observed during
this process. In this paper, we further reveal that physical quantities
associated with scalar hair exhibit identical oscillation patterns
during the evolution of the black hole-scalar field system. Moreover,
the oscillation frequency exhibits a linear dependence on the gauge
coupling constant of the scalar field with other parameters fixed.
Meanwhile, the horizon radius of hairy black holes and the mass within
the horizon increase monotonically with the gauge coupling constant.
We have also identified a critical initial charge value that distinguishes
hairy solutions that trigger black hole bombs from those that do not. 
\end{abstract}
\maketitle

\section{Introduction}

We know that the black hole is an object with an extremely strong
gravitational field, absorbing all matter, including light. However,
Penrose initially showed that the existence of an ergoregion allows
one to extract energy and angular momentum from a Kerr black hole
and to amplify energy via particle collisions \citep{Penrose:1969pc,Penrose:1971uk}.
Zel'dovich showed that dissipative rotating bodies amplify incident
waves \citep{ZelDovich1971JETP}. Teukolsky and Press soon quantitatively
discuss the superradiant scattering from a spinning black hole \citep{Press:1972zz,Press:1974zz}.
This work introduces the term ``superradiance'' in connection with
Zel'dovich's classical process of energy amplification. Superradiance
is the wave analog of the Penrose process: from particle energy extraction
to wave amplification \citep{ZelDovich1971JETP,Press:1972zz,Press:1974zz,Misner:1972kx,Bekenstein:1973mi}.
The amplified wave carries away energy, angular momentum, or charge
to infinity, and the black hole settles down to a black hole with
slightly smaller energy \citep{East:2014mfa,Baake:2016oku,Corelli:2021ikv,2025Hairless}.
However, if the amplified wave is reflected back towards the black
hole, the amplification can be repeated, leading to a run-away growth
of the wave, known as the ``black hole bomb'' \citep{Press:1972zz}.
The reflection could be realized by placing the black hole in an artificial
reflecting cavity or in anti-de Sitter spacetime \citep{Hawking:1999dp}.
Alternatively, for a massive bosonic field, the mass-term in the wave
equation naturally provides an effective reflecting potential barrier,
causing a black hole bomb without an artificial boundary \citep{Damour:1976kh}.
The mechanism by which the black hole bomb functions is called superradiant
instability, which operates at the full linear level. Considering
the backreaction of amplified waves on spacetime, the black hole bomb
will inevitably terminate at some point and the unstable seed black
hole must transition to a stable state, which is usually a hairy black
hole \citep{dias2011black,dias2012hairy,Hod:2012oa,Herdeiro:2014goa,Herdeiro:2016tmi}. 

A recently discovered mechanism triggers black hole bomb phenomena
by transforming stable RN black holes into hairy solutions, presenting
an alternative to the famous linear superradiant instability \citep{Zhang:2024aa}.
This approach, which triggers the transition through intrinsic scalar
field dynamics, contrasts with previous studies that relied on external
confinement---such as mirror-like boundary conditions \citep{Sanchis-Gual:2015lje,Sanchis-Gual:2016tcm,dias2018evading,ramon2021boson,davey2021phase}
or AdS boundary conditions \citep{dias2015black,Bosch:2016vcp,dias2017hairy,2017Collapse}---to
reflect and confine the scalar field. The model in \citep{Zhang:2024aa}
is built upon a massive, self-interacting complex scalar field minimally
coupled to the Maxwell field in general relativity. Within this framework,
the scalar field mass and self-interaction provide natural confinement,
operating without violating energy conditions to trigger the instability,
thereby eliminating the need for artificial mirrors or AdS boundaries.
The study reveals that after introducing an appropriate scalar field
into an RN black hole, strong nonlinear effects destroy the black
hole's stability. This causes the black hole to shed substantial energy
and charge, transitioning into a hairy black hole solution with reduced
horizon-enclosed energy, thereby effectively triggering a black hole
bomb. Unlike the dynamical evolution of superradiant instability,
which can be well understood from linear theory \citep{Bosch:2016vcp},
this process is intrinsically nonlinear and hard to analyze using
perturbation methods. 

During the evolution, the scalar hair undergoes prolonged, rhythmic
radial expansions and contractions. This behavior is analogous to
the radial pulsations observed in Cepheid variable stars and boson
stars. Cepheid variables are a type of pulsating variable star whose
outer atmosphere (and sometimes including deeper layers) undergoes
periodic expansion and contraction. These physical oscillations directly
cause the periodic variations in their luminosity and surface temperature
\citep{1912ap}. This period-luminosity relationship makes Cepheids
valuable ``standard candles'' in astronomy. By observing a Cepheid
variable's period, astronomers can determine its luminosity and then,
by comparing its apparent brightness (how bright it looks from Earth)
to its luminosity, calculate the distance to the star. This technique
has been instrumental in establishing the distances to nearby galaxies,
like the Magellanic Clouds, and even galaxies beyond our own Milky
Way \citep{Feast1999Cepheids,2008The,Beaton_2018,2024cepheids}. Correspondingly,
certain types of boson star solutions---in particular, scalar boson
stars---also exhibit a form of ``pulsation''. Their density profiles,
radii, or other properties oscillate periodically around an equilibrium
state over time. While this oscillation does not drive light variations
as in Cepheid variables, it constitutes an intrinsic dynamic behavior
\citep{1992bs,schunck2003general,Visinelli_2021,Liebling:2012fv}.
Returning to our investigation, such pulsations have been observed
for the first time in a black hole system. However, the temporal patterns
of these oscillations remain unclear. In this work, we investigate
in greater detail the oscillatory behavior of physical quantities
associated with scalar hair. We observe that these quantities exhibit
identical oscillation patterns. Furthermore, the oscillation frequency
is found to scale linearly with the gauge coupling constant of the
scalar field. 

The paper organizes as follows. In Section II, we introduce our general
model and list the basic equations we use to evolve a spherically
symmetric spacetime, the initial and boundary conditions, as well
as the numerical setup for the initial scalar perturbation. In Section
III, we conduct numerical simulations that illustrate in detail the
dynamical evolution of the RN black hole-scalar field system and the
resulting stable hairy black hole solution that forms after the evolution
stabilizes. In Section IV, we perform numerical simulations under
various initial parameters to analyze the dynamical oscillations of
the scalar field during the evolution. Finally, in Section V, we summarize
our results and discuss some possible future directions. 

\section{Model }

We consider the Einstein-Maxwell gravity minimally coupled with a
self-interacting massive complex scalar $\psi$. The Lagrangian density
is
\begin{equation}
\mathcal{L}=R-F_{\mu\nu}F^{\mu\nu}-D^{\mu}\psi(D_{\mu}\psi)^{\ast}-V(\psi),
\end{equation}
where $R$ represents the Ricci scalar associated with metric $g_{\mu\nu}$.
The Maxwell field strength $F_{\mu\nu}=\partial_{\mu}A_{\nu}-\partial_{\nu}A_{\mu}$
with $A_{\mu}$ being the gauge potential. The gauge covariant derivative
$D_{\mu}=\nabla_{\mu}-iqA_{\mu}$ with $q$ the gauge coupling constant
of the complex scalar field $\psi$. We derive the following equations
of motion
\begin{equation}
R_{\mu\nu}-\frac{1}{2}g_{\mu\nu}R=T_{\mu\nu}^{\psi}+2T_{\mu\nu}^{A},
\end{equation}
\begin{equation}
\nabla_{\mu}F^{\mu\nu}-\frac{1}{4}iq(\psi^{*}D_{\mu}\psi-\psi(D_{\mu}\psi)^{*})g^{\mu\nu}=0,
\end{equation}
\begin{equation}
D^{\mu}D_{\mu}\psi-\frac{\partial V}{\partial|\psi|{}^{2}}\psi=0,
\end{equation}
where $R_{\mu\nu}$ is the Ricci tensor and $T_{\mu\nu}$ the energy-momentum
tensor of the matter fields. The model is invariant under a local
$U(1)$ gauge transformation $A_{\mu}\rightarrow A_{\mu}+\partial_{\mu}\chi,\psi\rightarrow\psi e^{iq\chi},$
where $\chi$ is a regular real function of spacetime coordinate.

The energy-momentum tensor has two different components. The first
one, associated to the scalar field reads
\begin{gather}
T_{\mu\nu}^{\psi}=\frac{1}{2}(D_{\mu}\psi)^{*}(D_{\nu}\psi)+\frac{1}{2}(D_{\mu}\psi)(D_{\nu}\psi)^{*}\nonumber \\
-\frac{1}{2}g_{\mu\nu}(D^{\rho}\psi(D_{\rho}\psi)^{*}+V),
\end{gather}
while, the second one, associated to the Maxwell field is
\begin{equation}
T_{\mu\nu}^{A}=F_{\mu\rho}F_{\nu}^{\hspace*{0.3em}\rho}-\frac{1}{4}g_{\mu\nu}F_{\rho\sigma}F^{\rho\sigma}.
\end{equation}

We focus on the potential used in Q-ball literature \citep{Coleman:1985ki,Frieman:1988ut,Lee:1991ax,Kusenko:2001vu,Liebling:2012fv}:
\begin{equation}
V(\psi)=\mu^{2}|\psi|^{2}-\lambda|\psi|^{4}+\nu|\psi|^{6},\label{eq:V}
\end{equation}
with $\mu$ the scalar field mass, and $\lambda,\nu$ the positive
parameters governing the self-interactions of the scalar field. For
simplicity, hereafter we choose $\mu=1,\lambda=200,\nu=10000$, defining
the potential as $V(\psi)=|\psi|^{2}\left(1-\frac{|\psi|^{2}}{0.1^{2}}\right)^{2},$
which satisfies the weak and dominant energy conditions.

For dynamic simulation in spherically symmetric spacetime, we use
the spherical Painlevé-Gullstrand (PG) coordinates:
\begin{equation}
ds^{2}=-(1-\zeta^{2})\alpha^{2}dt^{2}+2\alpha\zeta dtdr+dr^{2}+r^{2}d\Omega^{2},\label{eq:ds2}
\end{equation}
where $\alpha,\zeta$ are metric functions dependent on $t$ and $r$.
This coordinate system is regular at the apparent horizon $r_{h}$
where $\zeta(t,r_{h})=1$. Taking the gauge potential $A_{\mu}dx^{\mu}=Adt$
and introducing auxiliary variables
\begin{equation}
\Phi=\partial_{r}\psi,\label{eq:drPhi}
\end{equation}
\begin{equation}
\Pi=\frac{1}{\alpha}(\partial_{t}\psi-iqA\psi)-\zeta\Phi,\label{eq:dtPhi}
\end{equation}
\begin{equation}
B=\frac{1}{\alpha}\partial_{r}A.\label{eq:B/dAt}
\end{equation}
The Einstein equations can be reduced to
\begin{equation}
0=\partial_{r}\alpha+\frac{\alpha r\text{Re}(\Pi^{*}\partial_{r}\psi)}{2\zeta},\label{eq:dra}
\end{equation}
\begin{equation}
0=\partial_{r}\zeta+\frac{\zeta}{2r}-\frac{r(\Pi\Pi^{*}+\Phi\Phi^{*}+2B^{2}+V)}{4\zeta}-\frac{r\text{Re}(\Pi\Phi^{*})}{2},\label{eq:drzeta}
\end{equation}
\begin{equation}
0=\partial_{t}\zeta-\frac{\alpha r}{2}[\Pi\Pi^{*}+\Phi\Phi^{*}+(\zeta+\frac{1}{\zeta})\text{Re}(\Pi\Phi^{*})].\label{eq:dtzeta}
\end{equation}
The Maxwell equations give 
\begin{equation}
0=\partial_{r}B+\frac{2B}{r}+\frac{q}{2}\text{Im}(\Pi\psi^{*}),\label{eq:dBr}
\end{equation}
\begin{equation}
0=\partial_{t}B-\frac{q}{2}\alpha\text{Im}[(\zeta\Pi+\Phi)\psi^{*}].\label{eq:dBt}
\end{equation}
The scalar equation becomes 
\begin{equation}
0=\partial_{t}\Pi-\frac{\partial_{r}[\alpha(\zeta\Pi+\Phi)r^{2}]}{r^{2}}-iqA\Pi+\alpha\psi\frac{\partial V}{\partial|\psi|{}^{2}}.\label{eq:dtPi}
\end{equation}
Given the initial scalar field distributions $\psi_{0},\Phi_{0}$
and $\Pi_{0},$ we first solve the constraint equations (\ref{eq:dBr},\ref{eq:drzeta},\ref{eq:dra},\ref{eq:B/dAt})
to obtain the initial values for the metric fields $B_{0},\zeta_{0},\alpha_{0}$
and $A_{0}.$ This completes the data on the initial time slice. Subsequently,
we use the Runge-Kutta method to solve the evolution equations (\ref{eq:dtPhi},\ref{eq:dtPi})
to calculate $\psi$ and $\Pi$ on the next time slice, while solving
the constraint equations (\ref{eq:drPhi},\ref{eq:dBr},\ref{eq:drzeta},\ref{eq:dra},\ref{eq:B/dAt})
to determine $\Phi,B,\zeta,\alpha$ and $A$ on the corresponding
time slice. By repeating this procedure, metric and matter field data
for all time slices are obtained.

\subsection{Physical quantities}

We monitor the evolution of several quantities during our analysis.
The scalar field energy is calculated as follows:
\begin{equation}
E_{\psi}=\frac{1}{4\pi}\int_{r_{h}}^{\infty}dV\rho_{\psi},
\end{equation}
where scalar field energy density $\rho_{\psi}=T_{\mu\nu}^{\psi}n^{\mu}n^{\nu}=(|\Pi|^{2}+|\partial_{r}\psi|^{2}+V)/2,$
and $n^{\mu}=(\alpha^{-1},-\zeta,0,0)$ the unit normal vector to
the constant time slice. The black hole charge can be determined using
the formula
\begin{equation}
Q_{h}=\frac{1}{4\pi}\oint_{r_{h}}dSF^{\mu\nu}n_{\mu}s_{\nu}=-\left.r^{2}B\right|_{r_{h}},
\end{equation}
where the normal vector $n_{\mu}=(-\alpha,0,0,0),$ and $s_{\nu}=(0,1,0,0)$
is the outward pointing unit normal vector to the apparent horizon
\citep{Torres:2014fga}. The scalar field charge is associated with
the Nother charge for the conserved current $j_{\mu}$:
\begin{equation}
Q_{\psi}=\frac{1}{4\pi}\int_{r_{h}}^{\infty}dVn^{\mu}j_{\mu},
\end{equation}
where $j_{\mu}=-\frac{q}{2}\text{Im}(\psi^{*}D_{\mu}\psi)$ is the
conserved current. The total charge of the system $Q=Q_{h}+Q_{\psi}$
should remain unchanged during dynamical evolution.

In dynamical simulation, the Christodoulou-Ruffini formula \citep{Christodoulou:1971pcn}.
is often employed to monitor the evolution of black hole mass \citep{East:2014mfa,Baake:2016oku,Corelli:2021ikv,East:2017ovw,East:2018glu}.
In situations where the black hole spin is absent, it is defined as
\begin{equation}
M_{B}=M_{h}+\frac{Q_{h}^{2}}{4M_{h}},\label{eq:MB}
\end{equation}
where $M_{h}=\sqrt{S_{h}/16\pi}$ is the irreducible mass and $S_{h}=4\pi r_{h}^{2}$
the apparent horizon area.

\subsection{Initial and boundary conditions}

We choose the initial seed black hole to be an RN black hole with
$\alpha_{0}=1$ and $\zeta_{0}(r)=\sqrt{\frac{2M_{0}}{r}-\frac{Q_{0}^{2}}{r^{2}}}.$
Here, $M_{0}=1$ and $Q_{0}=0.9M_{0}$ represent the mass and charge
of the black hole, respectively, and the parameter $M_{0}$ serves
to establish the energy scale. Thus the initial black hole mass $M_{B}=1$
and the total charge $Q=Q_{0}=0.9M_{0}.$ Solving $1-\zeta_{0}(r)^{2}=0$
yields the initial outer horizon radius of the RN black hole as $r_{h0}=1.43589.$
Then we impose following ingoing initial scalar pulse on the seed
RN black hole: 
\begin{equation}
\delta\psi=0.1pe^{-(\frac{r-12M_{0}}{2M_{0}})^{2}},\ \delta\Pi=\partial_{r}\delta\psi.\label{eq:IScalar}
\end{equation}
The initial pulse carries no net charge since its current $\delta j_{\mu}$
is vanishing everywhere. The total charge is unchanged by the pulse,
while the total mass increases with amplitude $p$. To ensure consistency
with Einstein’s equations, we explicitly solve the constraint equations
(\ref{eq:B/dAt},\ref{eq:dra},\ref{eq:drzeta},\ref{eq:dBr}) on
each time slice during the numerical evolution. In addition, we compute
the Hamiltonian constraint residual $H=\partial_{r}\zeta+$ (matter
contributions) and monitor that $H\approx0$ throughout the evolution,
which serves as a quantitative validation that our numerical solutions
remain consistent with the Einstein constraint equations.

Equation (\ref{eq:dra}) determines the evolution of $\alpha,$ with
boundary conditions given by 
\begin{equation}
\left.\alpha\right|_{r\rightarrow\infty}=1.
\end{equation}
This particular choice of boundary condition stems from the auxiliary
freedom of $\alpha dt$ in PG coordinates and implies that an observer
at infinity will measure time using the proper time coordinate $t$.
And at spatial infinity, the spacetime must asymptotically approach
Minkowski spacetime, requiring the solution to satisfy the following
additional boundary conditions: 
\begin{equation}
\left.\psi,\Pi,\Phi\right|_{r\rightarrow\infty}=0,\ \left.A\right|_{r\rightarrow\infty}=\frac{Q}{r},\ \left.B\right|_{r\rightarrow\infty}=-\frac{Q}{r^{2}}.
\end{equation}
They imply that $\left.\partial_{t}\psi\right|_{r\rightarrow\infty}=\left.\partial_{t}\Pi\right|_{r\rightarrow\infty}=0$
in (\ref{eq:dtPhi},\ref{eq:dtPi}). These conditions are sensible
as matter cannot reach spatial infinity in a finite time.

As previously mentioned, we use (\ref{eq:drzeta}) to solve for the
initial values of $\zeta$. For this purpose, we specify the following
boundary condition: 
\begin{equation}
\left.\zeta\right|_{t=0,r=r_{c}}=\zeta_{0}(r_{c}),
\end{equation}
where $\zeta_{0}(r)$ refers to the metric function of the initial
seed RN black hole, and $r_{c}=0.99r_{h0}$ denoting a cutoff located
close to the initial apparent horizon $r_{h0}=1.43589$ from the interior.
At asymptotic spatial infinity, we have 
\begin{equation}
\zeta=\sqrt{\frac{2M}{r}}(1+O(1/r)),
\end{equation}
where $M$ is the total mass of the system, accounting for the energy
of the gravity, Maxwell and scalar fields \citep{1996Gravitational}.
Furthermore, since the Arnowitt-Deser-Misner (ADM) mass in PG coordinates
is invariably zero and fails to reflect the correct physical mass
of the spacetime \citep{2016Numerical}, we employ the Misner-Sharp
(MS) mass $m(t,r)=\frac{r}{2}\zeta(t,r)^{2}$ to quantify the total
mass of the system \citep{MR177783}:
\begin{equation}
M=\underset{r\rightarrow\infty}{\lim}\frac{r}{2}\zeta(t,r)^{2}.
\end{equation}

\section{dynamic evolution of scalar fields}

We monitor key physical variables throughout the evolution, including
the scalar field value $|\psi|$ and charge $Q_{\psi},$ the scalar
field energy $E_{\psi}$ and energy density $\rho_{\psi},$ along
with black hole charge $Q_{h}$ and irreducible mass $M_{h}.$ While
charge and energy are transferred from and to the black hole, we confirm
that the charge conservation $Q_{h}+Q_{\psi}=Q$ holds rigorously.
Our studies reveal that RN black holes remain stable under small perturbations.
Only significantly large perturbations can drive nonlinear instabilities---triggering
black hole bomb---which ultimately induces a transition to hairy
black holes. This mechanism differs from the linear superradiant instability,
which drives unstable black holes to hairy states under specific boundary
conditions. Without such confinement (e.g., reflective boundaries
or an AdS spacetime), the scalar field is radiated away instead of
condensing into a stable hairy configuration.

To gain deeper insights, we impose an initial scalar field perturbation
of amplitude $p=0.26$ in numerical simulations. Varying the gauge
coupling constant $q$ reveals that hairy black hole solutions emerge
exclusively within the range $qM_{0}\in[2.998,5.286]$. For smaller
$q,$ the black hole initially exchanges energy and charge with the
scalar field, but rapidly absorbs the scalar field entirely, resulting
in an enlarged bald RN black hole. For larger $q,$ the evolution
resembles the weak-coupling case albeit with more pronounced transient
energy-charge exchange. Ultimately, all extracted energy and charge
are reabsorbed by the black hole, forming a more massive bald RN black
hole.

\begin{figure}[h]
\begin{centering}
\includegraphics[width=0.99\linewidth]{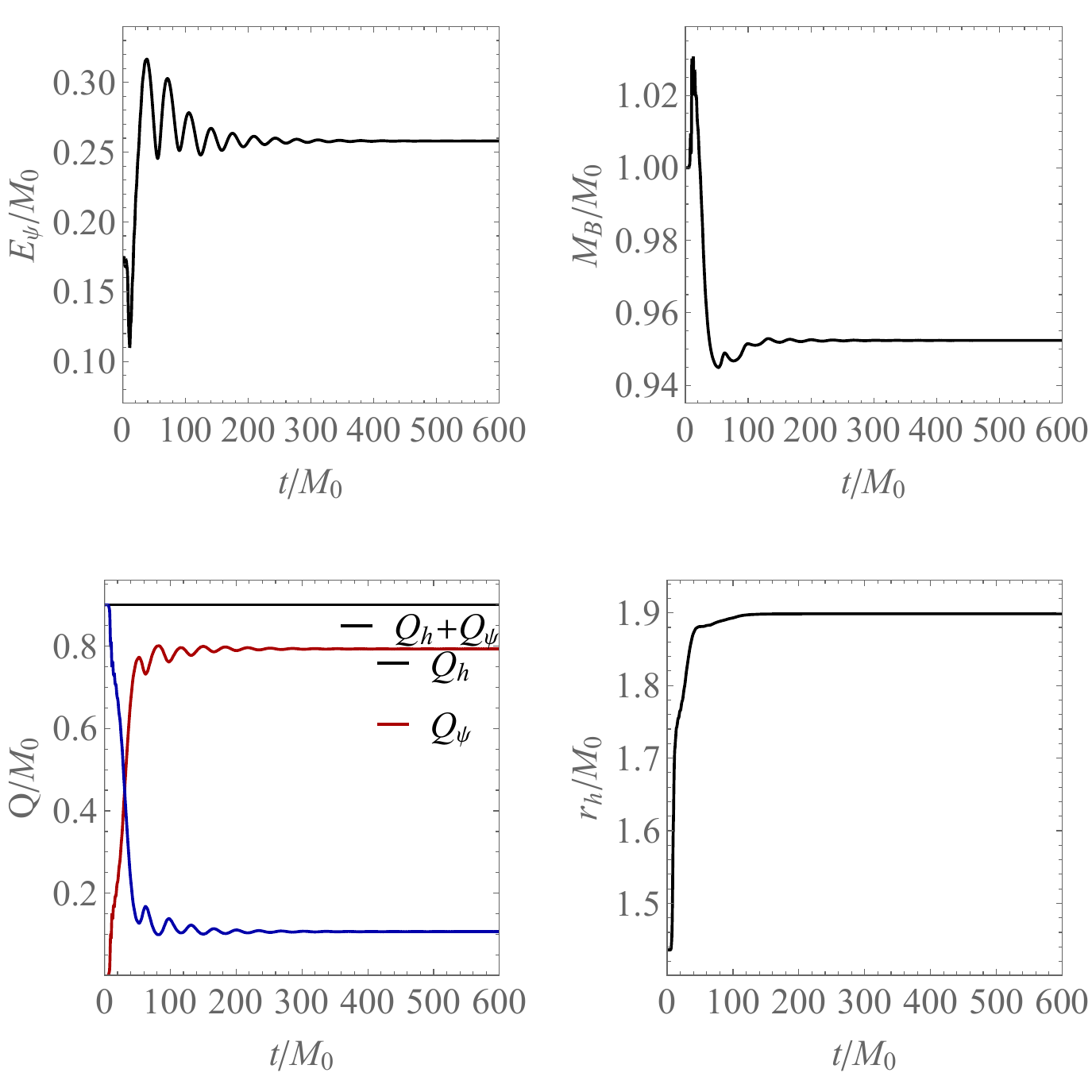}
\par\end{centering}
{\footnotesize\caption{{\footnotesize\label{fig1}The evolution of the scalar field energy
$E_{\psi},$ black hole mass $M_{B},$ scalar field charge $Q_{\psi},$
black hole charge $Q_{h}$ and total charge $Q$ starting from an
RN black hole with $M_{0}=1,Q=0.9M_{0}$ under the perturbation (}\ref{eq:IScalar}{\footnotesize )
with} $p=0.26$ when $qM_{0}=4$.}
}{\footnotesize\par}
\end{figure}

Without loss of generality, we further fix $qM_{0}=4$ to simulate
the evolution and present the dynamical results. In Figure \ref{fig1},
we observe that the scalar field energy $E_{\psi},$ black hole mass
$M_{B},$ scalar field charge $Q_{\psi}$ and black hole charge $Q_{h}$
all experience periodic growths and declines over a long period. These
quantities oscillate with identical frequencies, a feature that will
be analyzed in detail in subsequent sections. Periodic variation of
scalar field energy, similar to the nonlinear dynamics observed in
Kerr black holes \citep{Arvanitaki:2010sy,Yoshino:2015nsa} and confined
RN black holes \citep{Sanchis-Gual:2015lje,Sanchis-Gual:2016tcm,Bosch:2016vcp},
arising from the nonlinear development of superradiant instability.
Although the superradiant instability is initially a linear effect,
its long-term evolution and eventual saturation require a fully nonlinear
treatment. The periodic variation in scalar field energy seen in our
simulations shares similar features with these nonlinear regimes,
where the system undergoes energy exchange between the black hole
and the scalar condensate before reaching equilibrium. The total charge
$Q=0.9M_{0}$ remains unchanged, but a significant amount of charge
transfers from the black hole to the scalar field. Meanwhile, the
event horizon radius $r_{h}$ increases, indicating that the black
hole's horizon area $S_{h}$ exceeds that of the seed black hole.
This confirms the thermodynamic admissibility of the nonlinear process,
as it satisfies the second law of black hole thermodynamics---the
non-decreasing area law.

\begin{figure}[h]
\begin{centering}
\includegraphics[width=0.96\linewidth]{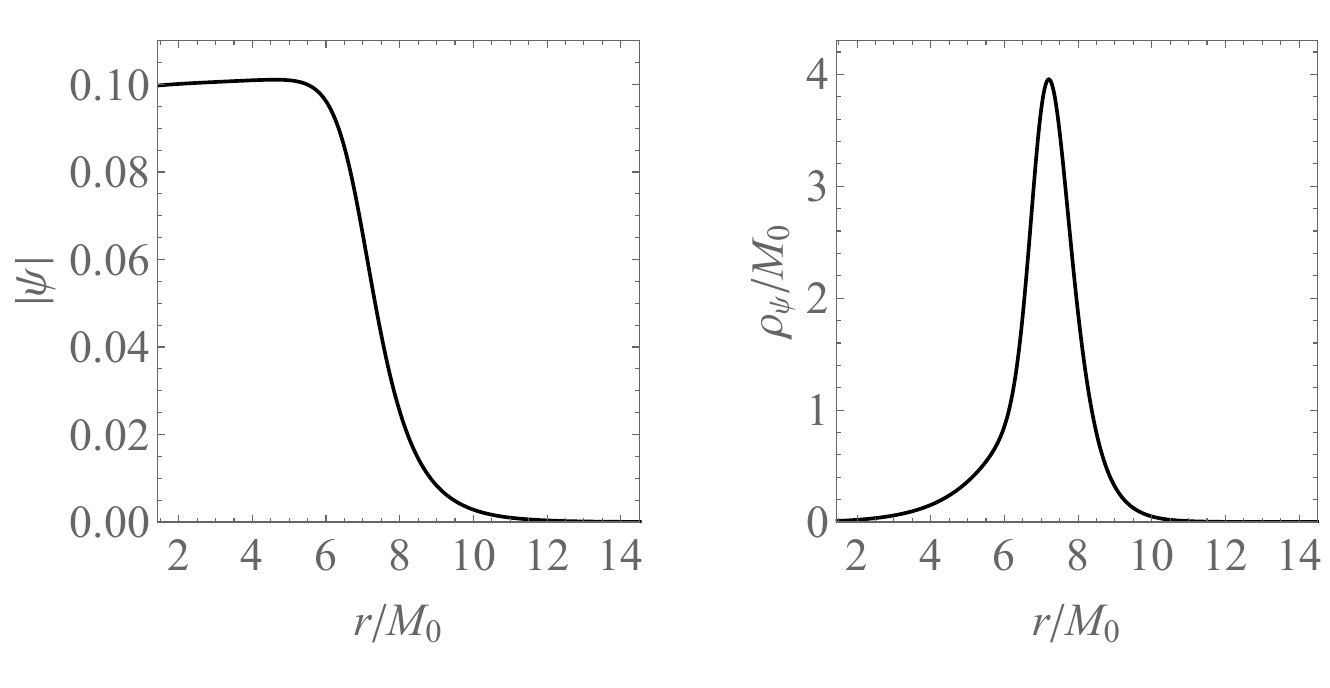}
\par\end{centering}
{\footnotesize\caption{{\footnotesize\label{fig2}The} ultimate stable distributions of the
scalar field value $|\psi|$ and of the energy density $\rho_{\psi}.${\footnotesize{}
Initial parameters: $M=M_{0},Q=0.9M_{0},p=0.26,qM_{0}=4$.}}
}{\footnotesize\par}
\end{figure}

The end-state after prolonged oscillatory evolution is a stable hairy
black hole with mass $M_{B}<M_{0}$ within the horizon, as shown in
Panel 2 of Figure \ref{fig1}. This implies that the initial RN black
hole of mass $M_{B}=M_{0}$ releases substantial energy outward during
its transition to a hairy state, exhibiting an explosive behavior
that resembles the dynamics of a black hole bomb. However, it is important
to clarify that in this context, we are not invoking the classical
superradiant instability typically associated with black hole bombs.
Instead, the term ``black hole bomb'' refers to the significant
energy release during the nonlinear evolution of the scalar field,
which drives the transition from a stable black hole to a hairy black
hole. This explosive energy release is a characteristic feature of
our model, but it differs from the runaway instability seen in traditional
black hole bomb systems. In Figure \ref{fig2}, we present the ultimate
stable distributions of the scalar field value $|\psi|$ and of the
energy density $\rho_{\psi}.$ Near the black hole, the scalar field
is confined to the vacuum state with $|\psi|=0.1$. Moving away from
the black hole, the scalar field gradually transitions to the vacuum
state with $|\psi|=0$. The scalar field energy density exhibits a
wave packet profile similar to the initial pulse configuration in
equation (\ref{eq:IScalar}), but it still oscillates laterally (radial
oscillations) with an amplitude too minuscule to be observable.

\begin{figure}[h]
\begin{centering}
\includegraphics[width=0.96\linewidth]{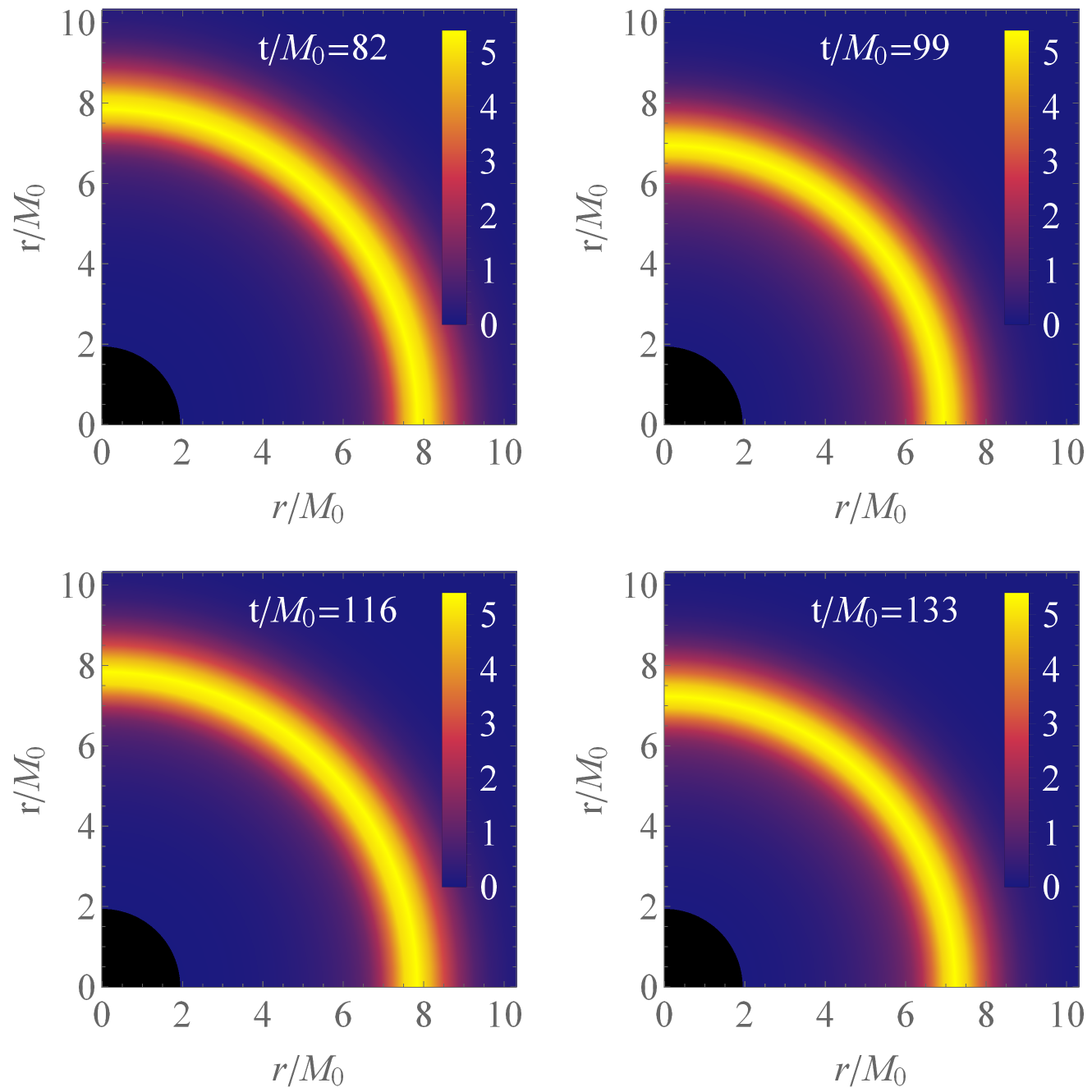}
\par\end{centering}
{\footnotesize\caption{{\footnotesize\label{fig3}The snapshots of the }energy density $\rho_{\psi}${\footnotesize{}
of the scalar field during the evolution. Showing one and a half oscillation
}period{\footnotesize s, Panels 1 and 3 correspond to maxima, while
Panels 2 and 4 correspond to a minima. The black regions in the lower
left corner represent the region inside the apparent horizon at the
corresponding times. Initial parameters: $M=M_{0},Q=0.9M_{0},p=0.26,qM_{0}=4$.}}
}{\footnotesize\par}
\end{figure}

Before reaching the final $Q$-hairy black hole solution, a fascinating
phenomenon emerges: the scalar field surrounding the black hole undergoes
prolonged periodic expansion-contraction cycles during evolution \citep{Zhang:2024aa}.
The time evolution of scalar field energy density $\rho_{\psi}$ also
corroborates this oscillatory behavior, as illustrated in Figure \ref{fig3}.
The $\rho_{\psi}$ manifests as a wavepacket undergoing radial oscillations,
with its temporal pattern corresponding precisely to observed periods
of energy-charge extraction and restoration of the scalar field, as
demonstrated in Figure \ref{fig1}. During each period, a small amount
of the scalar field energy and charge is absorbed by the central black
hole. This manifests as a progressive decay of peak values in the
energy density profile following each full period. After reaching
equilibrium, a stable $Q$-hairy black hole forms, in which the scalar
field exhibits persistent oscillations while its energy-momentum tensor
and geometry remain static, as evidenced in Figure \ref{fig1} and
Figure \ref{fig2}. In this work, the final state we obtain is consistent
with the static $Q$-hairy black hole solutions constructed in \citep{Hong:2020miv,Herdeiro:2020xmb},
supporting them as one of the possible equilibrium states. 

\section{analysis of radial oscillation frequencies in scalar fields}

\begin{figure}[h]
\begin{centering}
\includegraphics[width=0.96\linewidth]{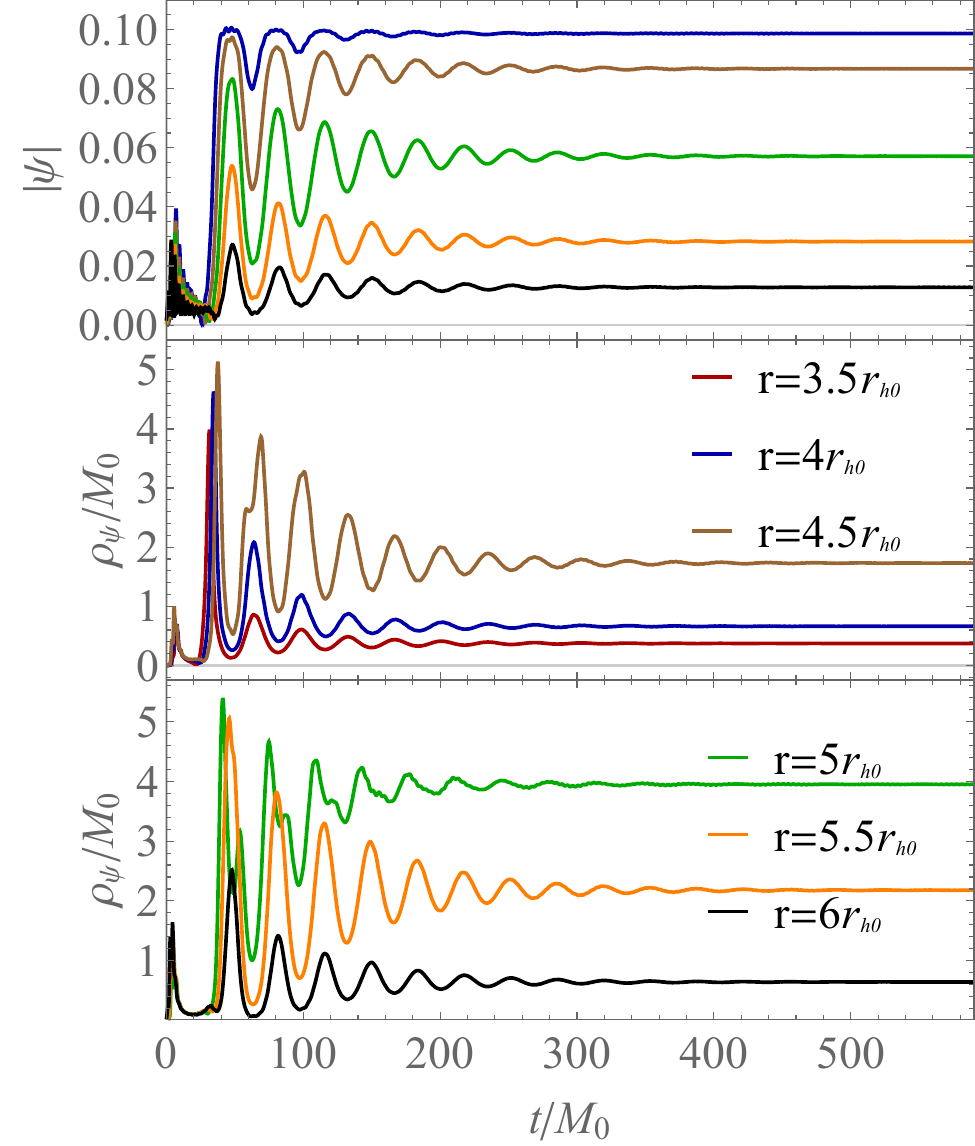}
\par\end{centering}
{\footnotesize\caption{{\footnotesize\label{fig4}Time evolution of the }scalar field value
$|\psi|$ and energy density $\rho_{\psi}${\footnotesize{} at radial
positions $r=3.5r_{h0},4r_{h0},4.5r_{h0},5r_{h0},5.5r_{h0},6r_{h0},$
where} $r_{h0}=1.43589$ {\footnotesize is the horizon radius. Initial
parameters: $M=M_{0},Q=0.9M_{0},p=0.26,qM_{0}=4.$}}
}{\footnotesize\par}
\end{figure}

As established in the preceding section, the evolution of scalar hair
exhibits the same temporal pattern in energy, charge, and energy density.
In this section, we provide a detailed analysis of this behavior.
Figure \ref{fig4} shows the evolution of scalar field value $|\psi|$
and energy density $\rho_{\psi}$ at several radial position points.
Following an initial phase of rapid dynamical variation, the scalar
field evolves into sustained periodic oscillations. During this process,
the scalar field value $|\psi|$ at radial positions $r=4r_{h0},4.5r_{h0},5r_{h0},5.5r_{h0},6r_{h0}$
exhibit identical oscillation frequency $\omega=\omega_{R}+i\omega_{I}=0.1846+0.0141i$.
The figure \ref{fig4} definitively confirms temporal alignment of
oscillation extrema across all monitored radial positions, with peak
and trough events exhibiting strict cycle-to-cycle correspondence.
Similarly, the energy density oscillations of the scalar field at
different radial positions are consistent and share the same oscillation
frequency $\omega$ ($0.1846+0.0141i$) as the variations of $|\psi|$.
The real part $\omega_{R}$ corresponds to the oscillatory behavior
of the scalar field energy, and the imaginary part $\omega_{I}$ controls
the decay of the field's amplitude. While the energy density of the
scalar field exhibits oscillatory decay, as shown in Figure \ref{fig4},
the spacetime itself does not exhibit significant time dependence
in the later stages of the simulation. This is because the decay of
the scalar field's amplitude leads to a gradual stabilization of the
system, and the black hole reaches a steady state, which gives the
appearance of a static spacetime. However, in the early stages of
evolution, the time-dependent oscillations of the scalar field, influenced
by the imaginary part of $\omega$, are more pronounced, leading to
a dynamic, though gradually stabilizing, spacetime. Moreover, oscillations
at radial positions $r=3.5r_{h0},4r_{h0},4.5r_{h0}$ and $r=5r_{h0},5.5r_{h0},6r_{h0}$
exhibit a $\pi$-phase shift, as these positions respectively reside
on opposite sides of the energy density wavepacket peak in Panel 2
of Figure \ref{fig2}.

\begin{figure}[h]
\begin{centering}
\includegraphics[width=0.96\linewidth]{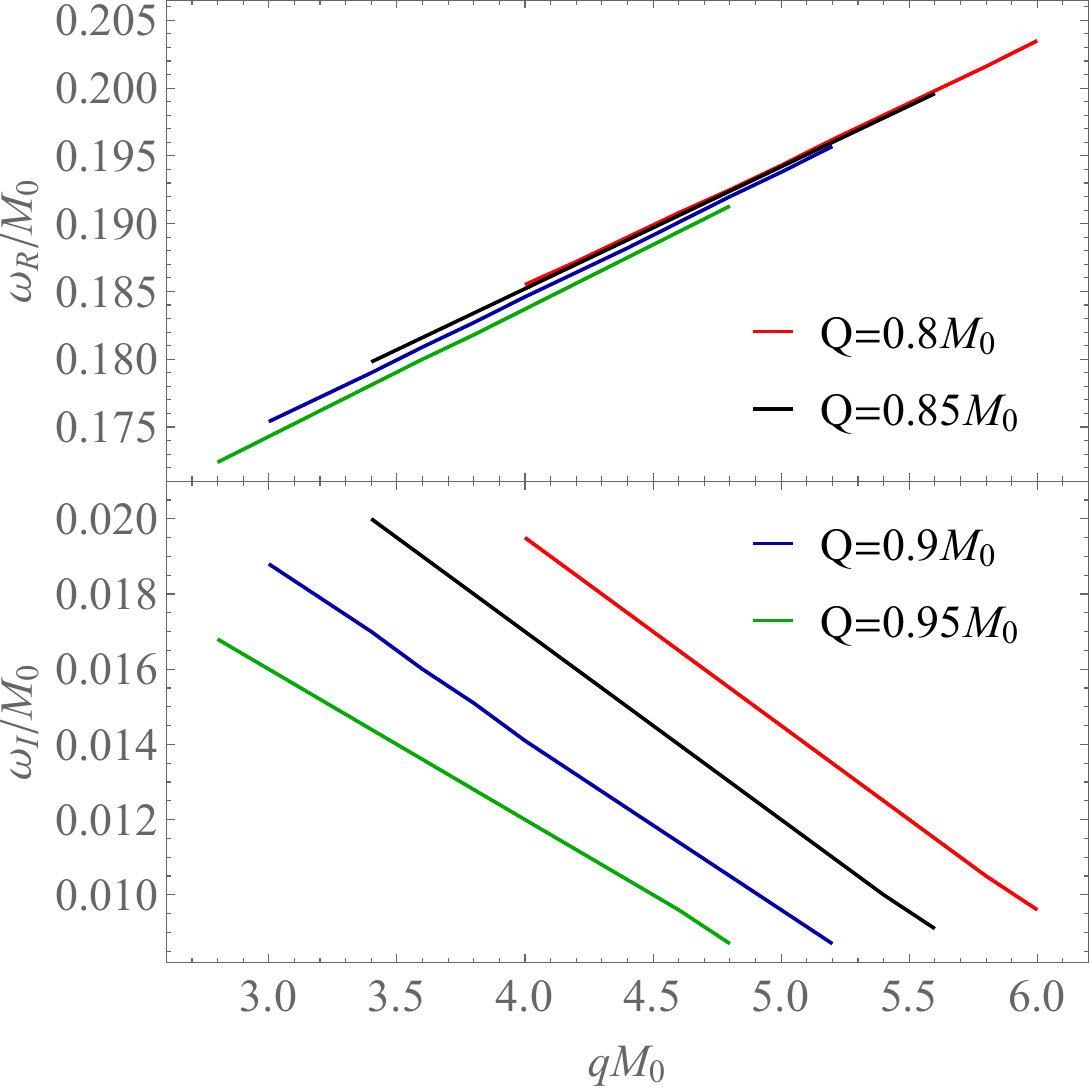}
\par\end{centering}
{\footnotesize\caption{{\footnotesize\label{fig5}Variation of the scalar field} oscillation
frequency $\omega_{R}$(real part) and decay coefficient $\omega_{I}$(imaginary
part) versus both initial charge $Q${\footnotesize{} and gauge coupling
}constant $q.${\footnotesize{} Other initial parameters: $M=M_{0},p=0.26$
(fixed).}}
}{\footnotesize\par}
\end{figure}

We perform numerical evolution simulations for a range of initial
charges $Q$ and gauge coupling constants $q,$ recording parameter
regions where hairy black hole solutions exist. Computing the scalar
field frequencies for distinct hairy solutions, we find that for fixed
initial charge $Q,$ both $\omega_{R}$ (oscillation frequency) and
$\omega_{I}$ (decay coefficient) scale linearly with the gauge coupling
parameter $q,$ as shown in Figure \ref{fig5}. This indicates that
larger $q$ values yield higher-frequency oscillations in scalar field
value $|\psi|,$ energy density $\rho_{\psi},$ and related scalar
field physical quantities, alongside slower decay of peak amplitudes.
Linear regression yields slopes of $0.009,0.009,0.0092,0.0094$ for
the oscillation frequency $\omega_{R},$ and $-0.00498,-0.00498,-0.00461,-0.00402$
for the decay coefficient $\omega_{I},$ as determined from the linear
fits in the figure. For a fixed gauge coupling constant $q,$ smaller
initial charge $Q$ corresponds to larger values of the decay coefficient
$\omega_{I}$. This indicates faster decay of peak amplitudes in scalar
field quantities---including field value and energy density.

\begin{figure}[h]
\begin{centering}
\includegraphics[width=0.96\linewidth]{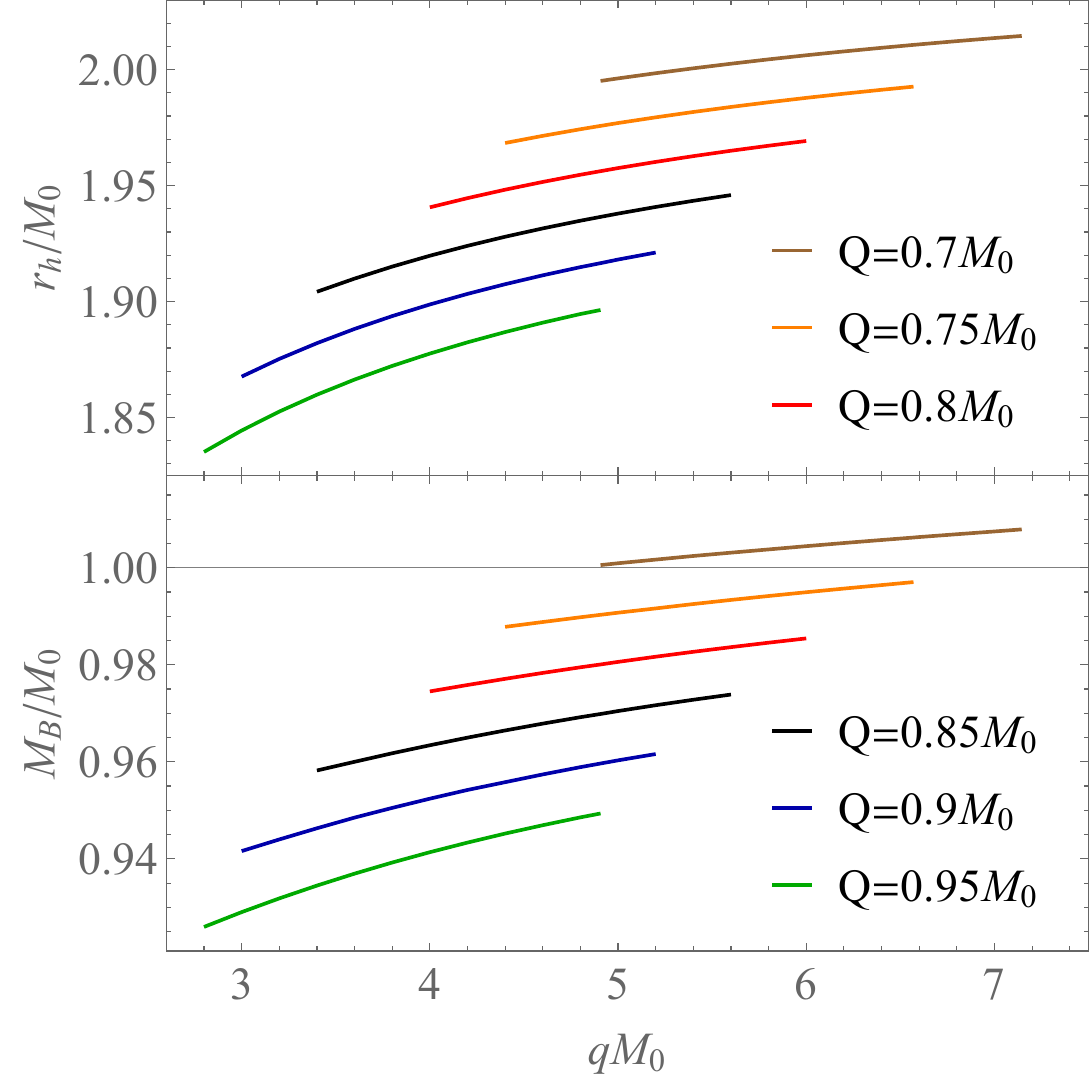}
\par\end{centering}
{\footnotesize\caption{{\footnotesize\label{fig6}Variation of the }event horizon radius
$r_{h}$ and black hole mass $M_{B}$ versus both initial charge $Q${\footnotesize{}
and gauge coupling }constant $q.${\footnotesize{} Other initial parameters:
$M=M_{0},p=0.26$ (fixed).}}
}{\footnotesize\par}
\end{figure}

Figure \ref{fig6} presents variations of horizon radius $r_{h}$
and black hole mass $M_{B}$ for hairy black hole solutions as functions
of initial charge $Q$ and gauge coupling constant $q.$ The figure
demonstrates that both smaller initial charge $Q$ and larger gauge
coupling constant $q$ yield hairy solutions with greater horizon
radius $r_{h}$ and black hole mass $M_{B}.$ Moreover, all obtained
hairy solutions exhibit horizon radii greater than those of the initial
seed black hole, demonstrating strict adherence to the non-decreasing
area law. For initial charges $Q=0.75M_{0},0.8M_{0},0.85M_{0},0.9M_{0},0.95M_{0}$
and varying gauge coupling constants $q,$ all resultant hairy black
hole solutions exhibit a black hole mass $M_{B}<M_{0}$ (where $M_{0}=1$),
demonstrating that the black hole bomb mechanisms are universally
triggered in these parametric configurations. The smaller $M_{B}$
is, the more energy the scalar field extracts from the black hole,
and the more pronounced the black hole bomb effect becomes. At $Q=0.7M_{0},$
the minimal gauge coupling constant ($q=4.907$) permitting hairy
solutions yields a black hole mass $M_{B}>M_{0}.$ This indicates
that while hairy solutions exist at this $Q$-value, they cannot trigger
black hole bombs. Therefore, under our given scalar field pulse configuration
and relevant parameters, the initial charge $Q_{c}=0.7M_{0}$ can
serve as a critical threshold that distinguishes whether a black hole
bomb can be created. In the charge range $0.7M_{0}<Q<0.75M_{0},$
hairy solutions exist that both trigger and fail to trigger black
hole bombs. Crucially, for any given $Q$ in this range, there exists
a specific range of $q$ values that trigger black hole bombs.

\section{Conclusion}

Unlike traditional superradiant instability driving unstable black
holes to form black hole bombs, our approach initiates from a linearly
stable Reissner-Nordström (RN) seed black hole. By introducing a massive
complex scalar field with nonlinear self-interactions, we creat hairy
black holes or trigger black hole bombs. This entire process is intrinsically
nonlinear throughout, and the obtained event horizon radius $r_{h}$
of the hairy black hole exceeds that of the seed black hole, satisfying
the non-decreasing area law. The end-state black hole mass $M_{B}$
within the horizon may exceed $M_{0}$ (the black hole absorbs scalar
field energy) or fall below $M_{0}$ (energy extraction triggers a
black hole bomb). It is worth noting that the injection of a scalar
pulse and the associated parameters must lie within specific ranges
to activate the transition.

Throughout the evolution, we observe that physical quantities associated
with the scalar hair exhibit identical oscillation patterns characterized
by a common frequency $\omega.$ This synchronization is rigorously
verified through the temporal evolution of these quantities across
distinct radial locations. Holding the initial scalar pulse fixed
while varying the seed black hole charge $Q$ and gauge coupling constant
$q$, we record parameter sets admitting hairy black hole solutions.
Our analysis reveals a linear relationship between the frequency $\omega$
(including both $\omega_{R}$ and $\omega_{I}$) and $q$ for fixed
$Q:$ $\omega_{R}$ increases linearly with $q,$ while $\omega_{I}$
decreases linearly with $q.$ Simultaneously, we observe that both
the event horizon radius $r_{h}$ and black hole mass $M_{B}$ increase
monotonically with $q.$ For each parameter pair ($Q,q$), numerical
simulations must be performed individually, and sufficient data must
be accumulated to establish the linear relationship between oscillation
frequencies and parameters. This computational process is highly time-consuming.
If we apply perturbation theory, which is similar to the treatment
of radial oscillations and stability analysis in boson stars \citep{2001Radial,alcubierre2021linear,kain2021boson},
we can conduct perturbation analysis of the partial differential equations,
combined with numerical computations to gain deeper insights into
the underlying physical mechanisms. This will be addressed in future
work.

Beyond the RN black holes discussed, analogous dynamical phenomena
manifest in Kerr black hole systems \citep{Herdeiro:2014pka,Herdeiro:2015tia}.
These phenomena also manifest in black holes with other types of solitary
hair, such as Proca and axion hair \citep{Liebling:2012fv,richard2015superradiance}.
The mass and self-interaction of the matter fields play crucial roles
in naturally confining and sustaining stable hairy solutions in asymptotically
flat spacetime, eliminating the need for artificial mirrors or AdS
boundaries. Overall, this nonlinear mechanism implies that analogous
dynamical oscillations may occur in other black holes or extreme compact
objects---a promising direction for future research.

\section*{Acknowledgments}

This work was supported by the National Natural Science Foundation
of China (NSFC) under Grants No. 12375048 and No. 12305080, the Guangzhou
Science and Technology Project under Grant No. SL2023A04J00576, and
the startup funding at South China University of Technology. 

\bibliographystyle{apsrev4-2ideal}
\bibliography{BHQhair}
 
\end{document}